\documentclass{astlb}
\usepackage{graphicx}
\usepackage{floatrow}
\usepackage{longtable}

\usepackage[hyperfootnotes=false,draft]{hyperref}

\usepackage{amsmath,amssymb}

\def\ps1{\textit{Pan-STARRS1}}

\def\doubleline{\vskip 3pt\hrule \vskip 1.5pt \hrule \vskip 5pt}

\begin{document}

\journalinfo{2023}{49}{8}{431}[444]

\title{Optical Identification of Galaxy Clusters among SRG/eROSITA X-ray Sources Based on Photometric Redshift Estimates for Galaxies}

\author{I. A. Zaznobin\address{1,2}\email{zaznobin@cosmos.ru},
  R.~A.~Burenin\address{1,2},
  A.~V.~Meshcheryakov\address{1},
  M.~R.~Gilfanov\address{1,3},
  N.~S.~Lyskova\address{1},
  P.~S.~Medvedev\address{1},
  S.~Yu.~Sazonov\address{1},
  R.~A.~Sunyaev\address{1,3}
  \addresstext{1}{\it Space Research Institute, Russian Academy of Sciences, Moscow, Russia}
  \addresstext{2}{\it Sternberg Astronomical Institute, Moscow State University, Moscow, Russia}
  \addresstext{3}{\it Max Planck Institut f\"{u}r Astrophysik, Garching, Germany}
  }
  
\submitted{}

\begin{abstract}  
We discuss an algorithm whereby the massive galaxy clusters detected in the SRG/eROSITA all-sky survey are identified and their photometric redshifts are estimated. For this purpose, we use photometric galaxies redshift estimates and WISE forced photometry. We select a sample of 634 massive galaxy clusters from the Planck survey with known spectroscopic redshifts in the range $0.1 < z_{spec} < 0.6$ to estimate the algorithm operation quality. The accuracy of the photometric redshift estimates for this sample is $\delta z_{phot}/(1+z_{phot}) \approx 0.5$\%, the fraction of large deviations is 1.3\%. We show that these large deviations arise mainly from the projections of galaxy clusters or other large-scale structures at different redshifts in the X-ray source field. Measuring the galaxy clusters infrared (IR) luminosities allows one to estimate the reliability of the optical identification of the clusters detected in the SRG/eROSITA survey and to obtain an additional independent measurement of their total gravitational masses, $M_{500}$. We show that the galaxy clusters masses $M_{500}$ estimated from their IR luminosity measurements have an accuracy $\sigma_{\lg\,M_{500}} = 0.124$, comparable to the accuracy of the galaxy clusters mass estimation from their X-ray luminosities.


\end{abstract}

\section{INTRODUCTION}

The search for and optical identification of galaxy clusters are among the important problem of extragalactic astronomy. Studying the galaxy clusters properties allows the structure and evolution of both the galaxy clusters themselves and the Universe as a whole to be studied. Studying various galaxy clusters samples allows the parameters of the cosmological models of the Universe being used to be constrained by measuring the galaxy cluster mass function \citep{av09cosm,PSZcosm13,PSZ2cosm}. In addition, investigating large galaxy clusters samples allows the baryon acoustic oscillations to be measured.

The Spectrum–Roentgen–Gamma (SRG) space observatory was launched on July 13, 2019 \citep{srg}. Two X-ray telescopes with grazing-incidence optics, eROSITA \citep{pred21} and ART-XC \citep{art} operating in the 0.2–-8 and 4--30~keV energy bands, respectively, are installed onboard the observatory. The study of galaxy clusters is one of the main tasks of the eROSITA survey. As of March 2022, four full all-sky surveys were completed. About 18~000 extended X-ray sources, most of which are identified with galaxy clusters, were detected in the eROSITA survey, in the sky region $0<|l|<180^\circ$, where the Russian scientists are responsible for processing the SRG/eROSITA all-sky survey data.

We organized the optical identification and spectroscopic redshift measurements of massive galaxy clusters from the eROSITA survey at optical telescopes \citep{zazn21,zazn21lh,br22hz,br23}. By now, the spectroscopic redshifts have been measured for more than 200 galaxy clusters with unmeasured spectroscopic redshifts, which in future will most likely enter into various cosmological samples of the SRG/eROSITA survey. Our observations will allow spectroscopic redshift measurements to be obtained for a cosmological sample of clusters with a size up to several thousand most massive galaxy clusters.

However, it is impossible to measure the spectroscopic redshifts for all of the galaxy clusters detected in the SRG/eROSITA survey, since this requires very large expenditure of observing time. Therefore, we set the goal to estimate the photometric redshifts of galaxy clusters using data from major photometric sky surveys. The most widely used method is to estimate the redshifts of galaxy clusters from the color of red-sequence galaxies \citep[see, e.g.,][]{redmapper,PSZ_RTT150}. There are a multitude of papers that provide the photometric redshifts of galaxies and galaxy clusters, including those obtained by machine learning methods \citep[see, e.g.,][]{mesh,zou,lsst,euclid,madpsz,redmapper,whl,zcluster}.

It is expected that using additional information about the exact positions and sizes of clusters, which can be obtained from X-ray data, will allow the accuracy and reliability of the photometric redshifts of galaxy clusters to be improved. In this paper we discuss a algorithm for the identification and estimation of the redshifts of massive galaxy clusters for which X-ray observations are available in the all-sky survey of the eROSITA telescope onboard the SRG space observatory. The photometric redshifts of galaxies from the catalog by \cite{zou} (hereafter HZ22) estimated by machine learning methods are used for the operation of the algorithm. We also use additional data from optical and infrared (IR) surveys. The distances are calculated in a standard way for the model of a flat Universe with parameters $H_0=70$~km\,s$^{-1}$\,Mpc$^{-1}$ and $\Omega_m = 0.3$. The algorithm was optimized for the most massive and richest clusters; for less rich clusters it must operate more poorly. We are going to study this question in our subsequent paper.

\section{DATA}

\subsection{The Sample of Galaxy Clusters}

To test the algorithm operation quality, we used a sample of massive galaxy clusters detected in the SRG/eROSITA all-sky survey in the sky area for the processing of which the Russian scientists are responsible ($0^{\circ} < l < 180^{\circ}$) and also detected in the Planck survey \citep{PSZ1,PSZ2,br17}. The spectroscopic redshifts of the galaxy clusters in the sample were taken from other catalogs \citep[see, e.g.,][]{160d,reflex,400d,mcxc,whl,PSZ1,PSZ2} and from the NASA/IPAC extragalactic database (\textit{NED}\footnote{https://ned.ipac.caltech.edu/}). We also used the spectroscopic redshifts of the galaxy clusters from the Planck survey taken from the papers of our group \citep{PSZ1,PSZ_RTT150,PSZ1Addendum,PSZ2,vorobyev16,br2018hz,zazn19,zazn20,zazn21,rttmos20,br22,br22hz,br23}.

During the first three SRG/eROSITA all-sky surveys a sample of 2317 galaxy clusters was obtained in the sky region for the processing of the SRG/eROSITA data of which the Russian scientists are responsible and which are also contained in the Planck survey. The spectroscopic redshifts were measured for 1053 of these clusters. The galaxy clusters at $z < 0.1$ have been well studied, while the galaxy clusters at high redshifts $z > 0.6$ have a significant scatter of photometric redshifts (see Table~3 from HZ22). Therefore, we optimized algorithm parameters to obtain as reliable identifications of the clusters as possible and as accurate photometric redshifts of the clusters located at $0.1 < z_{spec} < 0.6$ as possible. From this sample we selected 804 galaxy clusters at redshifts $0.1 < z_{spec} < 0.6$. Since we used the photometric redshifts of galaxies from DESI LIS \citep{desi} and WISE forced photometry \citep{wright10} for galaxies from the Pan-STARRS1 survey \citep{ps1}, the selection of objects was made in the region of overlap between DESI LIS and Pan-STARRS1. Therefore, the final sample of galaxy clusters used to estimate the operation quality of the optical identification and photometric redshift calculation algorithm contains 634 galaxy clusters at redshifts $0.1 < z_{spec} < 0.6$.

\begin{figure}
  \centering
    \includegraphics[width=1\columnwidth]{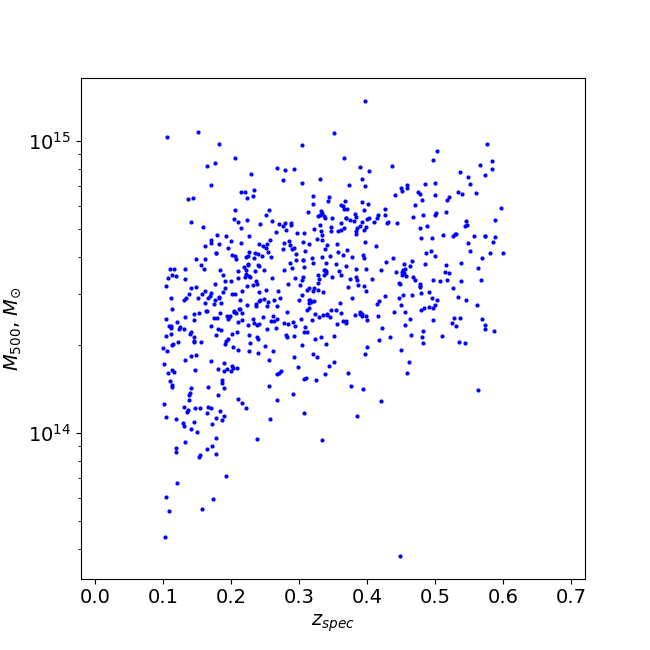}
    \caption{Relation between the clusters masses $M_{500}$ in our sample of 634 galaxy clusters from the SRG/eROSITA survey and their spectroscopic redshifts $z_{spec}$. The cluster masses were estimated from the eROSITA X-ray luminosities and calculated using the relation from \cite{av09xr}.
    }
  \label{m500_planck}
\end{figure}

The mass–redshift relation for the galaxy clusters in this sample of 634 galaxy clusters is presented in Fig.~\ref{m500_planck}; the masses were estimated in a standard method \citep{av09xr} from the X-ray luminosities derived from the eROSITA data. Almost all of the sample clusters are seen to be massive. Table~\ref{tab:cat_spec} shows the number of galaxy clusters in our sample encountered in other known catalogs of galaxy clusters, including those with spectroscopic redshift measurements. Table~1 separately shows the number of galaxy clusters whose spectroscopic redshifts were measured by our group, since there are spectroscopic measurements in different catalogs not for all galaxy clusters, the photometric redshifts are provided in many cases.

\begin{table*}
 \caption{The number of galaxy clusters in the sample encountered in other known catalogs.} 
  \label{tab:cat_spec}
  \renewcommand{\arraystretch}{1.1}
  \renewcommand{\tabcolsep}{0.35cm}
  \centering
  \footnotesize
  \begin{tabular}{lcc}
    \noalign{\doubleline}
    Source & Number of clusters & With measured $z_{spec}$\\
    \noalign{\hrule}
    Abell \citep{Abell} & 131 & 71\\
    PSZ2 \citep{PSZ2} & 148 & 126\\
    ACT \citep{act} & 129 & 98\\
    RM \citep{redmapper} & 486 & ---\\
    B17 \citep{br17} & 442 & 391\\
    WHL \citep{whl} & 422 & 164\\
    REFLEX \citep{reflex} & 68 & 68\\
    MCXC \citep{mcxc} & 93 & 93\\
    MACS \citep{macs} & 24 & 6\\
    Our group$^a$ & 121 & 121\\
    \noalign{\hrule}
  \end{tabular}
  
  \begin{flushleft}
  $^a$ -- The galaxy clusters in the sample for which our group measured the spectroscopic redshifts.\\
  \end{flushleft}
\end{table*}

\subsection{Galaxy Redshifts}

To determine the galaxy clusters redshifts, we used the photometric redshifts of galaxies presented in HZ22. This paper provides the galaxy redshifts estimated using data from the Dark Energy Spectroscopic Instrument (DESI), Dark Energy Survey (DES), and Hyper Suprime-Cam Subaru Strategic Program (HSC-SSP) sky surveys. The data are accessible for downloading from the site of the National Astronomical Data Center (NADC)\footnote{https://nadc.china-vo.org/} of the Ministry of Science and Technology of the People’s Republic of China.

In our work we used the photometric redshift estimates for galaxies based on data release 9 of the DESI Legacy Imaging Surveys \citep[DESI LIS,][]{desi}). This survey covers a large sky area compared to the DES and HSC-SSP surveys. In addition, this survey was conducted predominantly in the northern extragalactic sky in which the bulk of the galaxy clusters from the sample are located.

The DESI LIS data, data release 9, were published in January 2021 and consist of the photometric sky surveys performed at the Kitt Peak National Observatory in USA with the 2.3-m telescope in the \textit{gr} bands (Beijing-Arizona Sky Survey, BASS) and the 4-m telescope in the \textit{z} band (Mayall z-band Legacy Survey, MzLS) as well as with the 4-m telescope of the Cerro Tololo Inter-American Observatory, Chile, in the \textit{grz} bands (Dark Energy Camera Legacy Survey, DECaLS). The $5\sigma$ depth of the survey is $g = 24.7$, $r = 23.9$ and $z = 23.0$; the coverage is $\approx 20000$~deg$^2$ of the extragalactic sky $|b| > 18^{\circ}$ at declinations $-18^{\circ} < \delta < +84^{\circ}$.

In HZ22 the redshifts were determined optically extended objects brighter than $r < 23$ for the DESI LIS data. There are a total of $\approx 320$ million such sources in a sky area of 19876 deg$^2$. For 2.8 million galaxies in the fields of this survey their spectroscopic redshifts are known from open sources, such as the Sloan Digital Sky Survey \citep{sdssdr16} etc. The full list of spectroscopic surveys is given in Table~2 in \cite{zou19}.

The galaxies photometric redshifts were determined in HZ22 by constructing a local linear regression between the color space of the galaxies and their redshifts, where the galaxies were selected by the \textit{K}-nearest neighbor method. The method is based on the search for several galaxies with known redshifts with colors similar to those of the galaxies under study with unknown redshifts.

The quality of the galaxies photometric redshift determination is given in Table~3 in HZ22. It can be seen that for galaxies with $18 < r < 22$ at redshifts up to $z < 0.6$ the accuracy of the galaxies photometric redshifts from the DESI survey is better than the accuracy of the galaxies photometric redshifts from the DES and HSC-SSP surveys. The accuracy of the galaxies redshift measurements from DESI LIS is $\delta z/(1+z) = 0.0172$; the number of catastrophic outliers ($\delta z/(1+z) > 0.15$) 0.85\% of the total number.

For our work we took data for the catalog of galaxies photometric redshifts from the DESI survey. We selected only the galaxy coordinates given in the RA and DEC columns well as the photometric redshifts and their errors given in the PHOTO\_Z and PHOTO\_ZERR columns (below in the text photo\_z and photo\_zerr).

\subsection{X-ray Data}

We took the fluxes, coordinates, and sizes of the extended X-ray sources detected in the first three eROSITA all-sky surveys to identify the galaxy clusters and to estimate their redshifts. The sizes of the X-ray sources were determined as the core radius of the beta model $r_c$ by which the surface brightness profile of extended X-ray sources was fitted \citep{sarazin}.

The eROSITA data were processes using the eSASS (eROSITA Science Analysis Software System) package and the software developed by the group on the X-ray catalog of the eROSITA Russian Consortium. A more detailed description of the data processing and acquisition procedure is beyond the scope of this paper and will be discussed in subsequent works.

\subsection{IR Data}

We took WISE forced photometry \citep{br22} in the 3.4~$\mu$m band for the galaxies whose coordinates were determined from the Pan-STARRS1 data and were matched with the DESI catalog, data release~9. In \cite{br22} the photometry was obtained from the coadds with a longer exposure than that in DESI LIS; therefore, we took the photometry from \cite{br22}. These data are available for most of the galaxies from the HZ22 catalog. For the galaxies absent in the Pan-STARRS1 catalog we took the WISE photometry from DESI LIS.

\section{DESCRIPTION OF THE ALGORITHM}

The galaxy cluster redshifts are estimated by the algorithm divided into two steps. first, we perform the search for and identification of galaxy clusters. For this purpose, we calculated the IR luminosities of the galaxies near the central regions of the X-ray sources identified with galaxy clusters. Using the data of these galaxies from a small region near the cluster centers allows the cluster galaxies to be separated from the background and foreground galaxies.

Thereafter, the the galaxy clusters redshifts are determined from the galaxies photometric redshifts in a narrow redshift range at a larger radius comparable to $R_{500}$ or $R_{200}$. This technique allows one to estimate the cluster redshifts by eliminating the background galaxies or galaxy groups and to determine the the galaxy clusters redshifts more accurately from a large number of galaxies. Below, these steps will be described in detail.

\subsection{The Search for and Identification of Galaxy Clusters}

To search for galaxy clusters, we selected all galaxies from the HZ22 catalog at an angular distance from the X-ray source center no greater than five core radii of the beta model by which an extended X-ray source is described. At high redshifts the physical sizes of the covered regions of space can exceed manyfold the real sizes of the galaxy clusters. Therefore, we select only those galaxies for which the physical distance to the direction toward the X-ray source center did not exceed the typical size of massive clusters, approximately equal to 800 kpc --- the size $R_{500}$ of a galaxy cluster with a mass $M_{500} = 3 \cdot 10^{14} M_{\odot}$ at redshift $z = 0.6$. We used galaxies photometric redshifts from the HZ22 catalog to calculate the distances from the galaxies to the direction toward the X-ray source center.

Thereafter, we constructed the dependence of the galaxies IR luminosities on their assumed redshifts taken from HZ22. For this purpose, we calculated the luminosities of the galaxies in the WISE 3.4~$\mu$m band. The \textit{W1} forced photometry magnitudes from \cite{br22} are given in \textit{UBVRI} magnitudes (the Vega system). The flux was calculated using the effective width of the \textit{W1} band, $W_{eff} = 6626.42$\AA, and the reference spectral flux density in the \textit{UBVRI} system: $ZP_V = 8.1787\cdot 10^{-12}$~erg~s$^{-1}$~cm$^{-2}$~\AA$^{-1}$. The \textit{W1} magnitudes from HZ22 are given in the AB system; therefore, ZPV = $ZP_V = 9.5950\cdot 10^{-11}$~erg~s$^{-1}$~cm$^{-2}$~\AA$^{-1}$ for them.

\begin{displaymath}
   W1_{abs} = W1 - 5\lg D + 5
\end{displaymath}
\begin{displaymath}
   F_{W1} = ZP_V\cdot W_{eff}\cdot 10^{-0.4W1_{abs}}(1+photo\_z)/k_{cor}
\end{displaymath}
\begin{displaymath}
   L_{W1} = F_{W1}\cdot 4\pi\,(10pc)^2
\end{displaymath}

The K-corrections were calculated by shifting the template spectrum of an synthetic stellar population with an age of 2.5~Gyr and metallicity $Z = 0.008$ \citep{bc03} by the redshift photo\_z. To do this, we calculated the convolution of the template spectrum with the \textit{W1} passband profile. For this purpose, a Gaussian distribution was associated with each sample galaxy, where the expected value corresponds to the galaxy redshift photo\_z. The root-mean-square deviation was defined as the square root of the sum of the squares of the redshift measurement error photo\_zerr and the smoothing coefficient 0.02. The distribution was normalized to the luminosity in the 3.4~$\mu$m band:

\begin{displaymath}
   f(z) = \sum_i \frac{L_{W1i}}{\sqrt{2 \pi}\sigma_i}e^{-(z-photo\_z_i)^2/2\sigma_i^2}
\end{displaymath}
\begin{displaymath}
   \sigma_i = \sqrt{photo\_zerr_i^2+0.02^2}
\end{displaymath}

The distribution samples obtained for each galaxy were added and multiplied by the mean $<\sigma_{photo\_z}^i>\, = 0.0884$ in order that the maximum values of the IR luminosity distributions for the galaxies in the cluster field approximately match their real IR luminosities. The redshift at which the total IR luminosity of the galaxies in a given sky region is at a maximum will be deemed a preliminary cluster redshift estimate.

\subsection{Cluster Redshift Estimation}

The measurements of as many cluster galaxies as possible are required to be averaged to obtain a more accurate galaxy cluster redshift measurement. For this purpose, we selected galaxies in the cluster field at a great distance to the direction toward the X-ray source center, but in a narrow range of redshifts photo\_z. We selected galaxies with relative redshift errors $photo\_zerr/(1 + photo\_z) < 0.02$ with photo\_z in the range from $z - 0.06(1 + z)$ to $z + 0.06(1 + z)$, where z is the preliminary cluster redshift estimate. The galaxies were selected at a distance no greater than $1.4R_{500}$ to the direction toward the X-ray source center. Our estimates show that the number of selected background galaxies of the large-scale structures closest to the clusters located at redshifts adjacent to those of the galaxy clusters will increase with increasing distance, greater than this value. This will lead to a deterioration of the accuracy of the galaxy cluster redshift estimation.

We determined the values of $R_{500}$ based on the masses $M_{500}$ estimated from the X-ray luminosities of the galaxy clusters. We used the technique for calculating the X-ray luminosities of clusters from \cite{av09xr} given the K-corrections. When calculating the luminosities, we took into account the absorption of X-ray emission by galactic neutral hydrogen. The neutral hydrogen column densities were determined from \cite{nh}. The galaxy clusters redshifts were taken to be equal to the preliminary galaxy cluster redshift estimates.

For the selected galaxy clusters we calculated the weighted mean redshifts, where the galaxy weight was equal to the reciprocal of the square of its error photo\_zerr. From the samples of galaxies we excluded the galaxies whose photometric redshifts were outside two standard deviations of the galaxy redshift distributions. The galaxies were excluded from the samples until no galaxy was excluded in the next iteration. For the galaxy clusters from the Planck survey the photometric redshifts are estimated, on average, from the photometric redshifts of 50 galaxies.

\section{ESTIMATING THE RELIABILITY OF THE OPTICAL IDENTIFICATION}

Significant discrepancies between the estimated photometric redshifts of the galaxy clusters and their spectroscopic redshift measurements can arise during the operation of the algorithm. This is mainly because of the discrepancies in the optical identifications that arise predominantly when large-scale structures are projected onto extended X-ray sources. Some of the extended X-ray sources can also be formed by several closely spaced point sources and, hence, such sources should not be identified with galaxy clusters. Therefore, we propose the methods of estimating the reliability of the algorithm from the IR luminosities of galaxy clusters and from the IR luminosity distribution during the search for and identification of galaxy clusters.

\subsection{Distribution of the IR Luminosity in Redshift}
\label{sec:signi}

The value of the IR luminosity distributions maxima (peaks) can serve as a characteristic of the reliability of the galaxy cluster identification. For massive galaxy clusters the IR luminosity peaks must take high values exceeding the peaks that can arise when large-scale structures are projected in the fields of X-ray sources. In the case of projecting several clusters or large-scale structures, their X-ray fluxes can add up. Therefore, in some cases, it can be difficult to determine which galaxy clusters in the fields of X-ray sources are the most massive ones.

To estimate the reliability of the optical identification of galaxy clusters, we simulated a sample of 10~000 X-ray sources (below referred to as false sources) randomly distributed in the DESI LIS fields. The coordinates of the centers of these sources were selected in such a way that the angular distance between their centers was at least 10\arcmin. The fluxes and radii of the false sources were selected randomly among the galaxy clusters from the Planck survey detected in the SRG/eROSITA survey. In the false sources sample there must be virtually no objects located in the galaxy cluster fields. If such objects appear in the sample, then these coincidences are chance ones and reflect the probability that we will falsely identify the genuine X-ray source with the false one. The number of such chance coincidences is small, no more than a few tens.

\begin{figure}
  \centering
    \includegraphics[width=0.99\columnwidth]{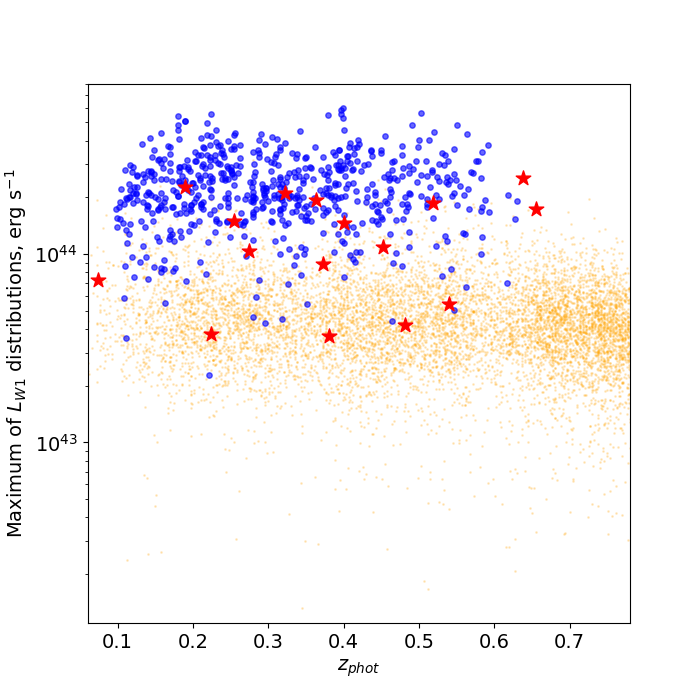}
    \caption{IR luminosity peaks of the clusters versus redshift. The blue dots are the clusters from the Planck survey, the orange dots are the false sources, and the red stars are the clusters from the Planck survey for which $(z_{phot} - z_{spec})/(1+z_{spec}) > 5\sigma$, where $z_{phot}$ are the preliminary galaxy cluster redshift estimates.
    }
  \label{1st_sign}
\end{figure}

In the field of each false source we obtained the IR luminosity distributions and determined the peaks and the redshifts at which they occur. The results of these simulations are presented in fig.~\ref{1st_sign}. This figure shows the IR luminosity peaks, where the orange and blue dots correspond to the false sources and the clusters from the Planck survey, respectively. The red stars correspond to the clusters from the Planck survey for which $(z_{phot} - z_{spec})/(1+z_{spec}) > 5\sigma$, where $z_{phot}$ are the preliminary cluster redshift estimates and $\sigma$ is the standard deviation.

The IR luminosity distributions for the clusters can contain more than one reliable local peak due to the correlation of background galaxies in redshifts in the fields of the galaxy clusters emerging in the projections of large-scale structures. In the cases of projections, it is often difficult to establish which galaxy clusters are more massive. The identification of galaxy clusters can be unreliable if in the field of their X-ray sources there are secondary peaks comparable to the main peak. In these cases, we can talk about the possible presence of projections in the fields of the X-ray sources if secondary peaks with an amplitude of more than half the main peak and a reliability greater than $2\sigma$ are observed in the IR luminosity distributions. Such cases can require additional data for the optical identification.

\subsection{The Mass–IR Luminosity Relation for Clusters}

The relation between the X-ray mass estimate for clusters and their IR luminosities can also be used to estimate the reliability of the galaxy clusters identifications. A correlation is known to be observed between the galaxy clusters IR luminosities and cluster masses \citep[see, e.g.,][]{lin,Kopylova,br17,br23}. The cluster masses can be estimated based on the eROSITA X-ray data \citep{av09xr} and their photometric redshifts. The IR luminosities of galaxy clusters can be estimated from the galaxies forced photometry in the 3.4~$\mu$m band.

To estimate the galaxy clusters IR luminosities, we selected galaxies for which the angular distances to the X-ray source centers did not exceed the angular sizes of $1.4R_{500}$. Among the galaxies samples obtained we selected galaxies whose photo\_z differed from the cluster photometric redshift by no more than two errors. No constraints were placed on the relative accuracy of the photometric redshifts for the galaxies. The galaxy clusters IR luminosities were calculated as the sum of the IR luminosities of all the selected galaxies including the K-correction. The luminosities and K-corrections of the galaxies were calculated by taking into account the photometric redshifts for each individual galaxy and not the entire cluster.

\begin{figure*}
  \centering
    \includegraphics[width=0.49\columnwidth]{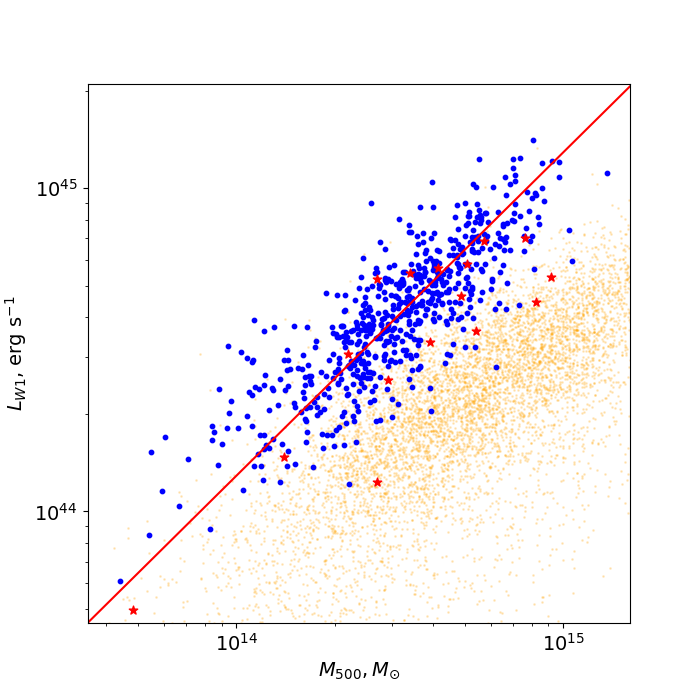}~\includegraphics[width=0.49\columnwidth]{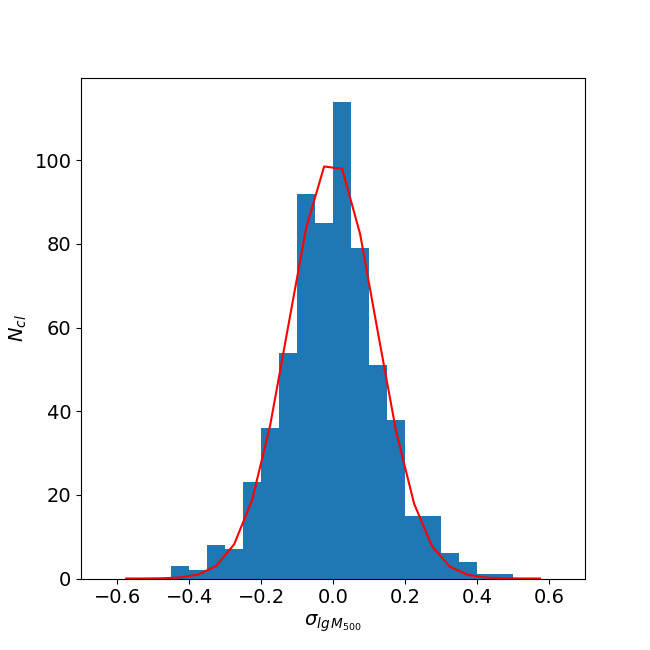}
    \caption{Left: luminosities $L_{W1}$ of the galaxy clusters in the 3.4~$\mu$m band versus their masses $M_{500}$. The blue dots are the galaxy clusters from the Planck survey sample, the red stars are the clusters from the Planck survey for which $(z_{phot} - z_{spec})/(1+z_{spec}) > 5\sigma$, and the orange dots are the false X-ray sources. The red line is the best fit $lg(L_{W1}) = lg(M_{500}) + k$. Right: a histogram of the distribution of $\sigma_{lg\,M_{500}}$ relative to the fit.
    }
  \label{lw1m500}
\end{figure*}

The results obtained are presented in fig.~\ref{lw1m500}. The left panel in this figure shows the galaxy clusters luminosities $L_{W1}$ in the 3.4~$\mu$m band as a function of their masses $M_{500}$. The red line indicates the best fit by a straight line $lg(L_{W1}) = lg(M_{500}) + k$, where $k = 30.11$, $L_{W1}$ is in erg~s$^{-1}$, and $M_{500}$ is in $M_{\odot}$. The blue dots are the galaxy clusters from the Planck survey sample, the orange dots are the false X-ray sources, and the red stars are the clusters from the Planck survey for which $(z_{phot} - z_{spec})/(1+z_{spec}) > 5\sigma$. It can be seen from the figure that the estimated IR luminosities for the false sources sample turn out to be severalfold lower than those for the massive clusters from the Planck survey sample. Consequently, low IR luminosities can point to unreliable identifications that can arise in the field of extended X-ray sources unidentified with galaxy clusters.

The right panel in Fig.~\ref{lw1m500} shows the distribution of cluster mass deviations. The red curve indicates the fit by a Gaussian. The standard deviation of the distribution is $\sigma_{lg\,M_{500}}^{tot} = 0.124$, i.e., $\pm 33$\%. The error of the mass determination for the clusters from their IR luminosities contains the errors in the cluster mass determination from the galaxy clusters X-ray luminosities. In \cite{av09xr} the deviation of the clusters X-ray luminosities relative to their masses is $\sigma_{ln\,L} = 0.396$, giving a deviation of the masses relative to the X-ray luminosities $\sigma_{lg\,M_{500}} = 0.107$. Consequently, the accuracy of the galaxy clusters mass estimation from their IR luminosities is comparable to the accuracy of the mass estimation from the X-ray luminosity.

\subsection{Determining the Reliability of the Optical Identification}

We defined the reliability of the optical identification as the product of the probability that the main peak of the IR luminosity distribution in redshift for the object under study is larger than the peaks of the false sources at a given redshift in the range $z = z_{phot}\, \pm \, 0.05$ and the probability that the ratio of the IR luminosity of the object under study to their mass $M_{500}$ estimated from the X-ray data is greater than the ratio of the IR luminosity to M500 for the false sources in the range $lg(M_{500})\, \pm \, lg2$.

As has already been mentioned previously, the IR luminosity distributions for the galaxy clusters can contain more than one reliable local peak due to the correlation of background galaxies in redshifts in the fields of the galaxy clusters emerging in the projections of large-scale structures. In the cases of projections, it can often be dificult to establish which galaxy clusters are more massive. The number of galaxy clusters from the Planck survey sample for which the IR luminosity distribution has secondary peaks whose probability exceeds $1\sigma$ and $2\sigma$ at the total number of 634 is 189 (29.8\%) and 22 (3.5\%), respectively.

The identification of galaxy clusters can be unreliable if there are secondary peaks comparable to the main peak in the field of their X-ray sources. In these cases, we can talk about the possible presence of projections in the fields of the X-ray sources if secondary peaks with an amplitude of more than half the main peak and a reliability greater than $2\sigma$ are observed in the IR luminosity distributions. In the cases of projections, the X-ray flux is the sum of two projected clusters. For this reason, the mass $M_{500}$ of the cluster located at the redshift of the peak in the IR luminosity distribution can be overestimated. Consequently, the IR luminosity of such a cluster will be underestimated compared to clusters of comparable mass and, hence, the reliability of such objects will be low.

Some point X-ray sources can be located at a small angular distance relative to one another and, for this reason, these sources are detected as one extended X-ray source. Such extended sources cannot be identified with massive galaxy clusters having high IR luminosities. A significant discrepancy between the IR luminosity and the values typical for galaxy clusters of similar mass can point to unreliability of the identification. Therefore, the reliability estimate for these objects will be lower than that for massive galaxy clusters.

\begin{figure*}
  \centering
    \includegraphics[width=0.49\columnwidth]{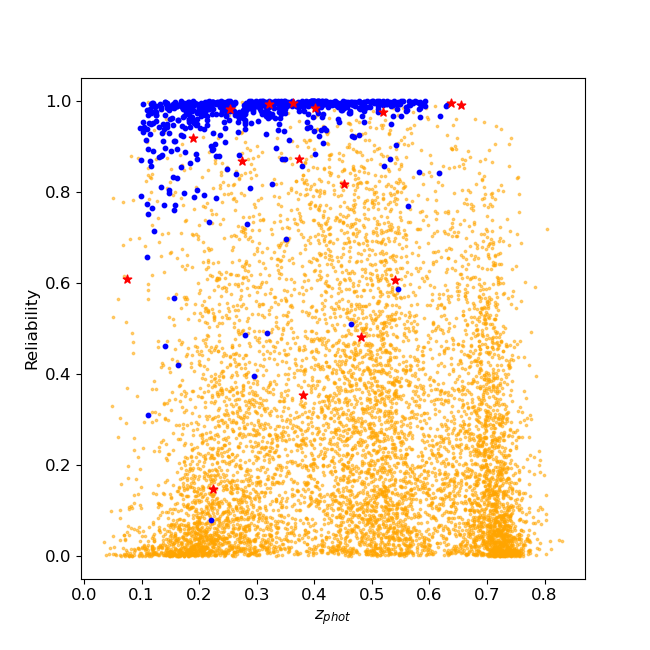}~\includegraphics[width=0.49\columnwidth]{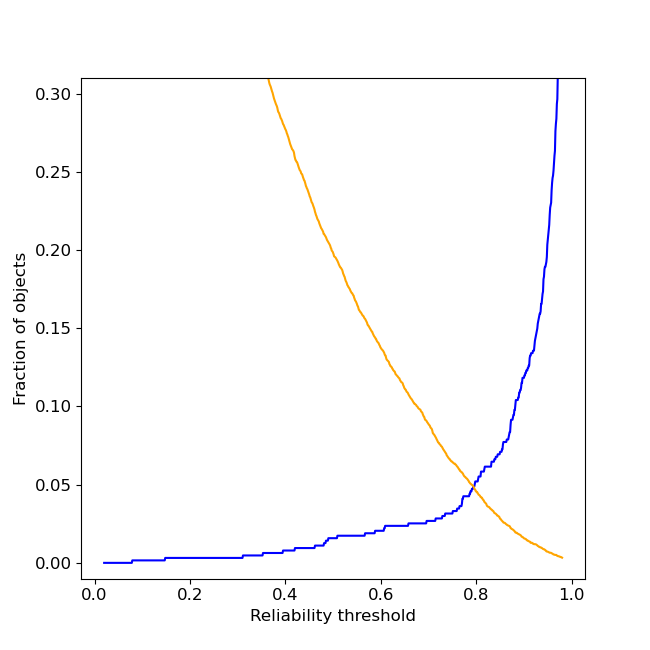}
    \caption{Left: the reliability estimate. The orange dots are the false clusters, the blue dots are the galaxy clusters from the Planck survey sample, and the red stars are the clusters from the Planck survey for which $(z_{phot} - z_{spec})/(1+z_{spec}) > 5\sigma$, where $z_{phot}$ are the preliminary galaxy cluster redshift estimates. The blue line on the right panel indicates the fraction of galaxy clusters from the Planck survey below a certain reliability threshold, the orange line indicates the fraction of false clusters above a certain reliability threshold.
    }
  \label{nadya}
\end{figure*}

The left panel in fig.~\ref{nadya} shows the reliability estimate. The orange dots are the false clusters, the blue dots are the galaxy clusters from the Planck survey sample, and the red stars are the clusters from the Planck survey for which $(z_{phot} - z_{spec})/(1+z_{spec}) > 5\sigma$, where $z_{phot}$ are the preliminary galaxy cluster redshift estimates. The reliability estimate for the massive clusters is seen to exceed considerably the reliability estimate for the false clusters. The blue line on the right panel indicates the fraction of galaxy clusters from the Planck survey below a certain reliability threshold, the orange line indicates the fraction of false clusters above a certain reliability threshold. Some galaxy clusters from the Planck survey sample are seen to have a low reliability.

The low reliability is related mainly to the contribution of active galactic nuclei or projected galaxy clusters located at small angular distances from the centers of extended X-ray sources to the X-ray flux. As a rule, this leads to an overestimation of the cluster masses and, therefore, the IR luminosities of the clusters will have values lower than the typical values for galaxy clusters of similar mass. In these cases, the IR luminosity peaks toward the X-ray sources can take values typical for massive galaxy clusters. Therefore, in these cases, we can talk about the possible contribution of other X-ray sources.

Another reason is the absence of photometric redshifts for some of the cD galaxies in the HZ22 catalog. This is probably related to the criteria for the selection of galaxies from DESI LIS in HZ22. There are several galaxy clusters from the Planck survey that have no photometric redshifts of cD galaxies in HZ22. Most of these clusters have a low reliability. The cD galaxies make a large contribution to the IR luminosities of the clusters and the IR distributions toward the cluster center and, therefore, the absence of photometric redshifts for cD galaxies can reduce the reliability of the galaxy clusters.

Setting a reliability threshold value will allow such objects to be found. If we set the threshold value to 0.8, then 5.4\% of the galaxy clusters from the Planck survey sample will have a reliability below this value. When the threshold value is reduced to 0.5, the fraction of galaxy clusters from the Planck survey sample below the threshold value will decrease to 1.7\%. The fraction of false clusters above the threshold values of 0.8 and 0.5 will be 4.6 and 19.8\%, respectively.

In the SRG/eROSITA all-sky survey the galaxy clusters are determined as extended X-ray sources fairly reliably. For example, if 10\% of the extended X-ray sources in the SRG/eROSITA survey are false, then for the reliability threshold values of 0.8 and 0.5 the number of false extended X-ray sources
incorrectly identified in the optical range as genuine galaxy clusters will be 0.5 and 2\% of the entire sample, respectively. It can be seen that by raising the reliability threshold, the number of objects falsely identified as galaxy clusters decreases. At the same time, the number of identified galaxy clusters, a significant fraction of which to a certain reliability threshold will be galaxy clusters with projections, will decrease.

\section{RESULTS}

\subsection{An Example of the algorithm Operation}

\begin{figure*}
  \centering
    \includegraphics[width=0.49\columnwidth]{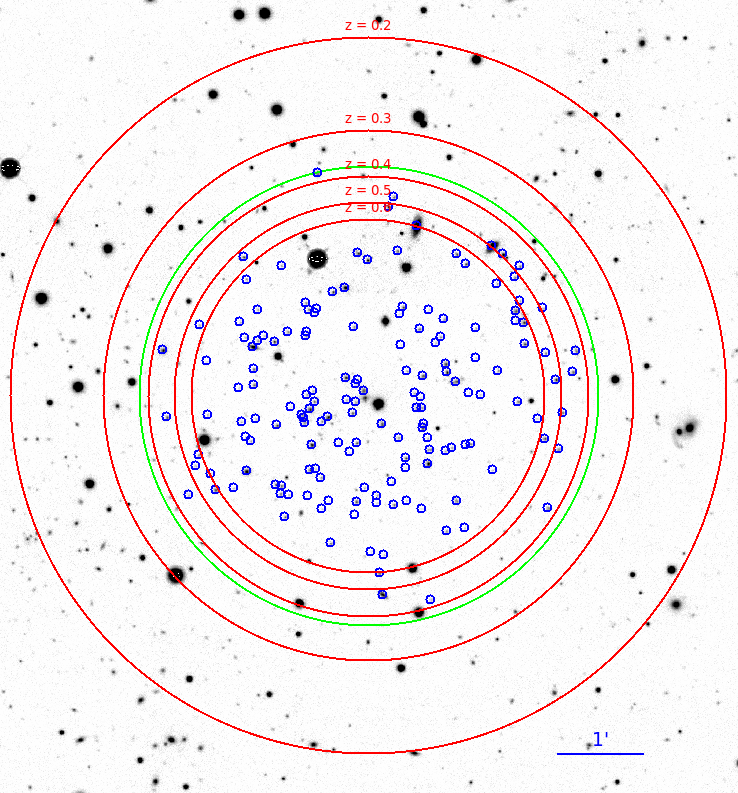}~\includegraphics[width=0.49\columnwidth]{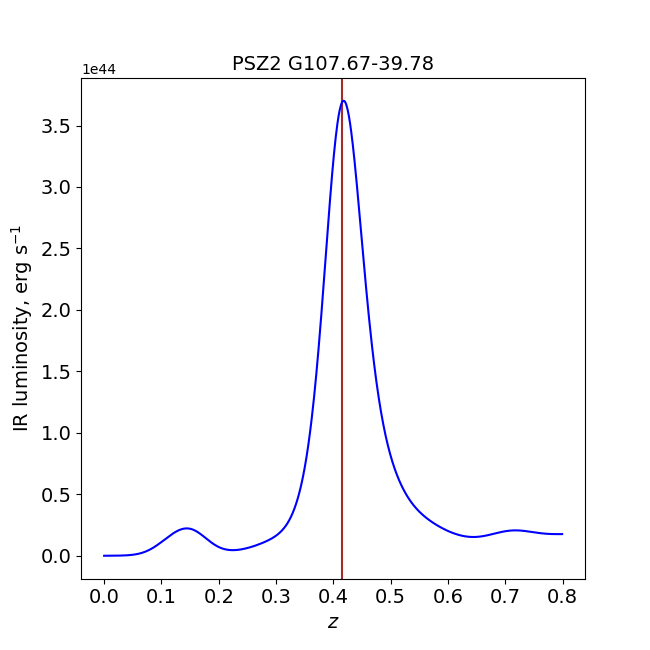}
    \caption{The galaxy cluster PSZ2\,G$107.67\!-\!39.78$. Left: the result of the selection of galaxies at the first stage for the cluster field, a DESI LIS image in the \textit{r} band, the image center corresponds to the eROSITA X-ray source center. The blue circumferences indicate the galaxies selected at the first stage. The red circumferences indicate the circles of radius 800~kpc at different distances. Right: the IR luminosity distribution in a cylinder in redshift. The vertical line marks the cluster spectroscopic redshift.
    }
  \label{1stage_ds9}
\end{figure*}

An example of the identification of a galaxy cluster from the IR luminosity peaks is shown in Fig.~\ref{1stage_ds9}. The left panel in this figure displays the field of the galaxy cluster PSZ2\,G$107.67\!-\!39.78$ from the second Planck catalog in the \textit{r} band from DESI LIS. The X-ray source core radius in this case is 31.7\arcsec. The blue circumferences indicate 156 galaxies selected by the algorithm at the galaxy cluster search and identification stage. The red circumferences mark the distance of 800~kpc to the direction toward the X-ray source center at different redshifts. The distribution of the IR luminosity in redshift is presented on the right panel in the figure; only one peak is seen. The preliminary redshift estimate is $z = 0.418$.

The X-ray source center coordinates are: 00~01~11.3 +21~32~14. The redshift measurements from SDSS, data release 13 \citep{sdssdr13}, are accessible for 13 galaxies of this cluster within 4\arcmin\ of the source center.

At the stage of the redshift determination for the galaxy cluster PSZ2\,G$107.67\!-\!39.78$ we selected 70 galaxies from which the cluster redshift was estimated. Our estimate of $z_{mean} = 0.4183$ agrees excellently with the preliminary redshift estimate of $z = 0.418$. The reliability is 0.998. The spectroscopic redshift of this galaxy cluster was determined as the mean of the spectroscopic redshift measurements for 13 cluster galaxies and is $z_{spec} = 0.4113 \pm 0.0025$. The error is 0.5\%: $(z_{phot} - z_{spec})/(1+z_{spec}) = 0.0049$.

\subsection{An Example of the Projection of Large-Scale Structures}

The above example shows the operation of the algorithm when there are no projections of large-scale structures, such as filaments, clusters, and galaxies groups, in the X-ray source field. In the overwhelming majority of galaxy clusters no projections of large-scale structures are observed. However, in the cases of the projections of galaxy groups and clusters at various redshifts located at a small angular distance from one another, the X-ray emission can add up into one X-ray source with a large angular size, as discussed above. Let us give an example of the operation of the algorithm in the case of the projection of two galaxy clusters at different redshifts identified with an X-ray source with a large angular size.

The projection of two clusters at redshifts $z = 0.1908$ \citep{zazn20,Stre} and $z = 0.3920$ \citep{zazn19} is observed in the field of the galaxy cluster PSZ2\,G$069.47\!-\!29.06$ from the second Planck catalog. It is pointed out in \cite{zazn19} that the cluster at $z = 0.3920$ is closer to the Sunyaev-Zeldovich source and is brighter in the infrared; therefore, the source PSZ2\,G$069.47\!-\!29.06$ was identified with the cluster at $z = 0.3920$.

\begin{figure*}
  \centering
    \includegraphics[width=0.49\columnwidth]{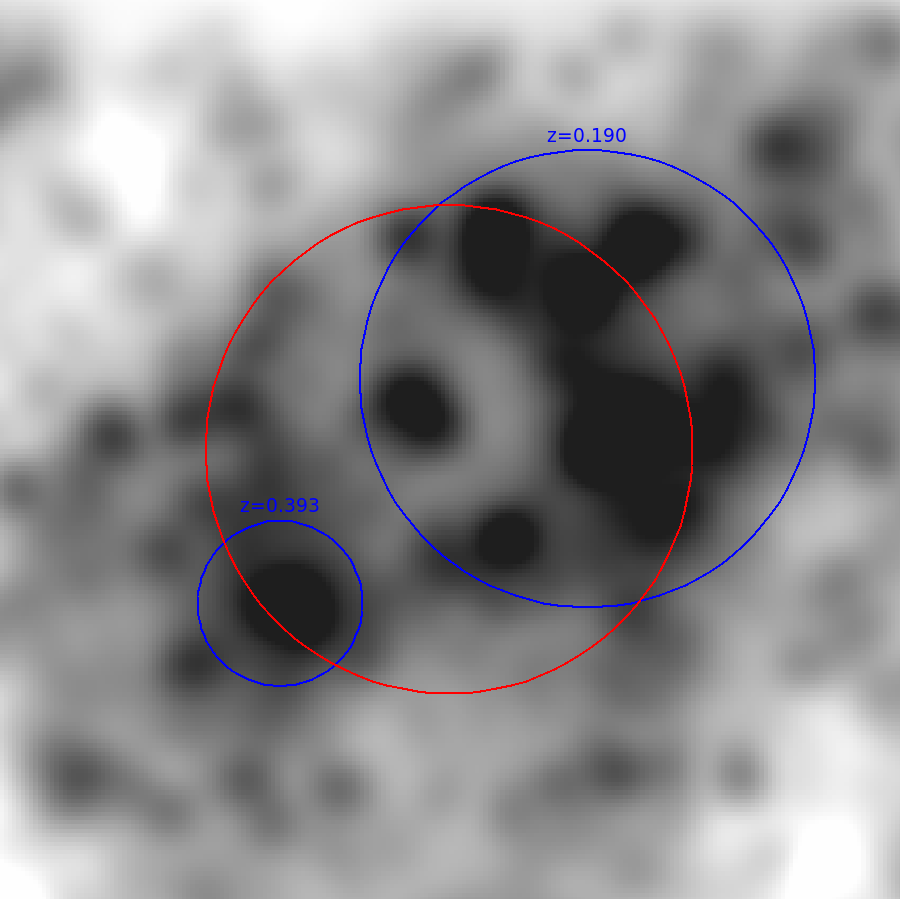}~\includegraphics[width=0.49\columnwidth]{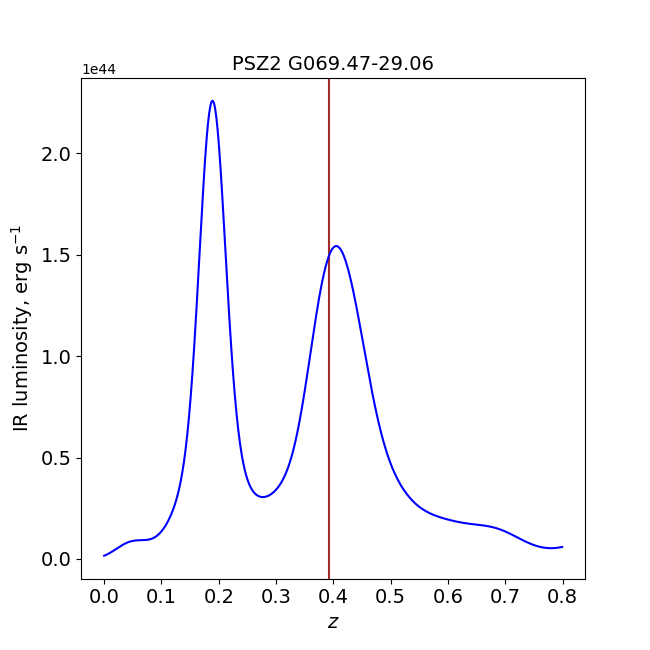}
    \caption{The cluster PSZ2\,G$069.47\!-\!29.06$. Left: the WISE image in the 3.4~$\mu$m band from \cite{zazn19} smoothed with a $\beta$-model of radius 24\arcsec. The red circle marks the eROSITA source of radius 158\arcsec\ identified with the cluster. Right: the IR luminosity distribution of galaxies in the central region of the cluster.
    }
  \label{G069}
\end{figure*}

The source PSZ2\,G$069.47\!-\!29.06$ in X-rays has a large radius $\approx 158$\arcsec. The left panel in Fig.~\ref{G069} shows a smooth image of the cluster field in the 3.4~$\mu$m band, as in Fig.~4 from \cite{zazn19}, where the red circumference indicates the X-ray source whose angular size is 158\arcsec. In this case, the galaxy clusters fluxes are seen to add up, which can increase the angular size of the X-ray source and its flux. The right panel in Fig.~\ref{G069} presents the IR luminosity distribution, where two peaks are seen. The red dots indicate the peaks with the preliminary redshift estimates of $z = 0.189$ and $z = 0.405$. The algorithm allows one to estimate this galaxy cluster redshift $z_{phot} = 0.1893$, which agrees very accurately with the spectroscopic measurements of $z_{spec} = 0.1908$ for one of the projection galaxy clusters. In this case, the reliability is 0.918.

\subsection{Photometric Redshift Estimates for Galaxy Clusters}

\begin{figure*}
  \centering
    \includegraphics[width=0.47\columnwidth]{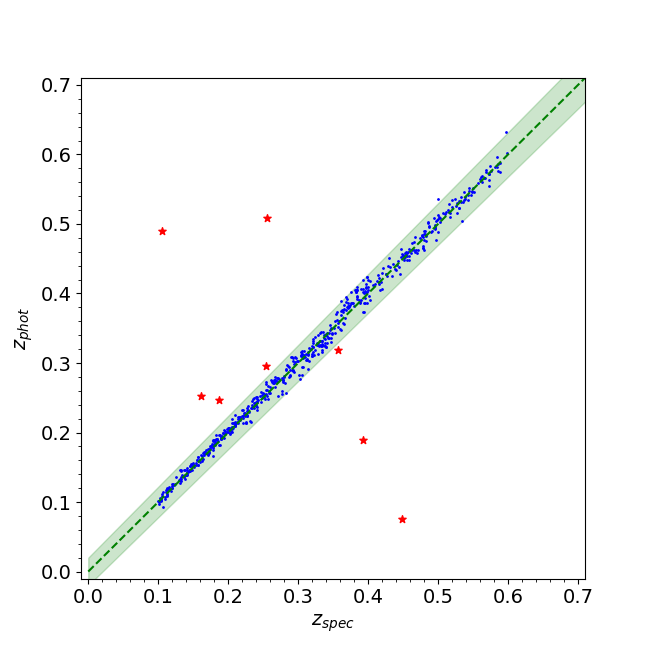}~\includegraphics[width=0.47\columnwidth]{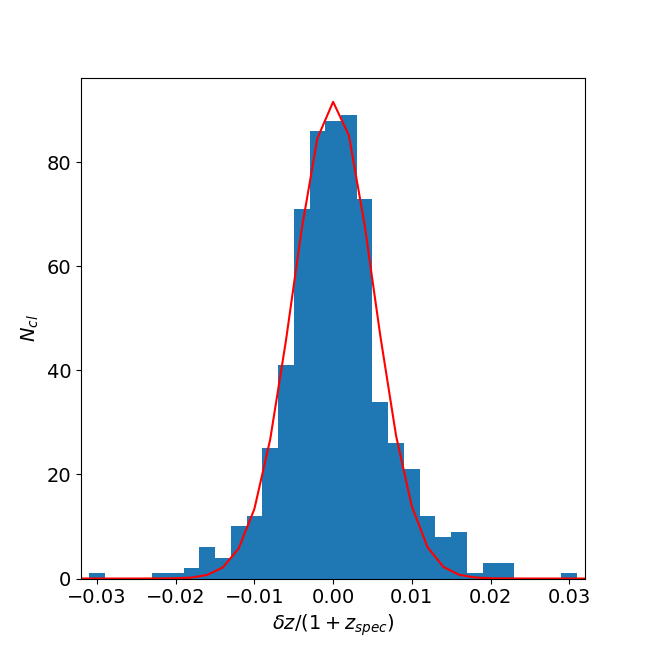}
    \caption{Photometric redshifts of the galaxy clusters. Left: a comparison of $z_{phot}$ with the spectroscopic $z_{spec}$ measurements. The red stars mark the galaxy clusters with possible projections; the green color designates the $4\sigma$ region. Right: a histogram of the galaxy clusters distribution in $(z_{phot}-z_{spec})/(1+z_{spec})$, the red line indicates the Gaussian fit to the distribution.
    }
  \label{fchart:cl}
\end{figure*}

In Fig.~\ref{fchart:cl} are shown the photometric redshifts for our sample of 634 galaxy clusters from the SRG/eROSITA survey at $0.1 < z_{spec} < 0.6$. On the left panel the photometric redshifts $z_{phot}$ obtained by our algorithm are compared with the spectroscopic $z_{spec}$ measurements. The green color designates the region where $(z_{phot}-z_{spec})/(1+z_{spec}) < 4\sigma$. The red stars indicate nine clusters for which photometric redshifts large deviations relative to the spectroscopic measurements due to the presence of projections are observed. In the fields of these galaxy clusters it can be difficult to distinguish the main cluster; therefore, differences in the identifications arise.

The right panel in Fig.~\ref{fchart:cl} presents a histogram $(z_{phot}-z_{spec})/(1+z_{spec})$ galaxy clusters distributions, the column width is 0.002, the red line indicates the Gaussian fit to the distribution. Table~\ref{tab:zphot} gives the parameters of this distribution both for the entire investigated redshift range $0.1 < z_{spec} < 0.6$ and for narrow ranges. In table are provided the clusters number, the shifts of the distribution centroid relative to $z_{phot} = z_{spec}$, and the standard deviation. The last column gives the galaxy clusters number with large photometric redshifts deviations relative to the spectroscopic redshift measurements from the catalog exceeding $5\sigma$, which are explained by the projections of galaxy clusters.

\begin{table}
  \caption{Results.} 
  \label{tab:zphot}
  \renewcommand{\arraystretch}{1.1}
  \renewcommand{\tabcolsep}{0.35cm}
  \centering
  \footnotesize
  \begin{tabular}{ccccc}
    \noalign{\doubleline}
    $z_{spec}^a$ & $N_{cl}^b$ & $\delta z_0^c$ & $\sigma^d$ & $>5\sigma^e$\\
    \noalign{\hrule}
    0.1 -- 0.2 & 159 & 0.0028 & 0.0038 & 3 (1.9\%)\\
    0.2 -- 0.3 & 158 & 0.0009 & 0.0047 & 2 (1.3\%)\\
    0.3 -- 0.4 & 162 & -0.0012 & 0.0091 & 2 (1.2\%)\\
    0.4 -- 0.5 & 94 & -0.0032 & 0.0058 & 1 (1.1\%)\\
    0.5 -- 0.6 & 61 & 0.0000 & 0.0044 & 0 (0\%)\\
    \noalign{\hrule}
    0.1 -- 0.6 & 634 & 0.0000 & 0.0051 & 8 (1.3\%)\\
    \noalign{\hrule}
  \end{tabular}
  
  \begin{flushleft}
  $^a$ -- The range of spectroscopic redshifts of the galaxy clusters.\\
  $^b$ -- The number of galaxy clusters.\\  
  $^c$ -- The value of $\sigma$ for the Gaussian fit of the $(z_{phot}-z_{spec})/(1+z_{spec})$ distribution.\\
  $^d$ -- The value of the Gaussian centroid.\\
  $^e$ -- The number of discrepancies between the photometric redshifts of the galaxy clusters and their spectroscopic measurements at the $5\sigma$ level arising from the possible projections.\\
  \end{flushleft}
\end{table}

The achieved accuracy of the galaxy clusters photometric redshifts for the exceeds the accuracy of the galaxies photometric redshifts from the HZ22 catalog in the investigated redshift range approximately by a factor of 3. The discrepancies number of between the galaxy clusters photometric redshifts and their spectroscopic measurements at the $5\sigma$ level arising from the possible projections is about 1.3\%\ in the entire investigated redshift range.

\section{CONCLUSIONS}

We described a simple algorithm for the optical identification of galaxy clusters from the SRG/eROSITA survey and the determination of their photometric redshifts in the redshift range $0.1 < z_{spec} < 0.6$ based on the galaxies photometric redshifts from DESI LIS using additional WISE forced photometry. An accuracy of the galaxy clusters photometric redshift determination of 0.5\% was achieved. The number of discrepancies between the galaxy clusters photometric redshifts and their spectroscopic measurements at the $5\sigma$ level is 1.3\%. Most of these discrepancies arise from the possible projections of large-scale structures in the X-ray sources fields.

We measured the galaxy clusters IR luminosities. The accuracy of the galaxy clusters masses estimating $M_{500}$ from their IR luminosities, $\sigma_{lgM_{500}} = 0.107$, be comparable to the accuracy of the galaxy clusters masses estimating $M_{500}$ from their X-ray luminosities. We estimated the reliability of the optical identification of galaxy clusters.

The achieved accuracy of the galaxy clusters photometric redshift determination, 0.5\%, is high compared to the accuracy of the estimation of their redshifts from the color of the red sequence, for example, the redMaPPer algorithm, whose accuracy at $z \approx 0.5$ is $(z_{phot}-z_{spec})/(1+z_{spec}) = 0.02$. Recently, \cite{madpsz} has achieved a comparable accuracy of 0.47\% for a galaxy clusters sample from the Planck survey using data from the SRG/eROSITA all-sky survey and DES. The accuracy of the photometric redshift estimates for clusters achieved in \cite{madpsz} is comparable to the accuracy in our work.

Note that the characteristics of the photometric redshifts obtained in our work refer to the most massive galaxy clusters. For less massive ones the reliability of their optical identification must be poorer. In future we will improve the algorithm for the identification of less massive galaxy clusters and the estimation of their redshifts. The sample of galaxy clusters from the 400 deg$^2$ survey of the ROSAT space X-ray observatory \citep{400d} is well suited for this purpose.

Our results are based on the photometric redshifts of DESI LIS galaxies from \cite{zou} estimated by machine learning methods. In future, our group will estimate the galaxies photometric redshifts based on photometric data from the Pan-STARRS1, DESI LIS, and WISE surveys using machine learning methods. It is expected that using a larger number of major optical and IR sky surveys will allow the accuracy of the photometric galaxies redshift determination to be improved. In turn, this will allow the accuracy of the galaxy clusters photometric redshifts to be improved.

\acknowledgements

This work was supported by RSF grant no. 21-12-00210. In this study we used observational data from the eROSITA telescope onboard the SRG observatory. The SRG observatory was built by Roskosmos in the interests of the Russian Academy of Sciences represented by its Space Research Institute within the framework of the Russian Federal Space Program, with the participation of the Deutsches Zentrum f\"{u}r Luft- und Raumfahrt (DLR). The SRG spacecraft was designed, built, launched, and is operated by the Lavochkin Association and its subcontractors. The science data are downlinked via the Deep Space Network Antennae in Bear Lakes, Ussurijsk, and Baykonur, funded by Roskosmos. The eROSITA X-ray telescope was built by a consortium of German Institutes led by MPE, and supported by DLR. The eROSITA data used in this work were processed using the eSASS software developed by the German eROSITA consortium and the proprietary data reduction and analysis software developed by the Russian eROSITA Consortium.


\begin{thebibliography}{90}


\bibitem[\protect\citeauthoryear{Abell \etal}{1989}]{Abell}
  \reference{Abell~G.~O., Corwin~H.~G.~Jr., Olowin~R.~P.}
  {\journal{\apjs}{\bf 70}{1}{1989}}

\bibitem[\protect\citeauthoryear{Böhringer \etal}{2004}]{reflex}
  \reference{Böhringer \etal}[Böhringer~H., Schuecker~P., Guzzo~L., Collins~C.~A., Voges~W. \etal]
  {\journal{\aap}{\bf 425}{367}{2004}}
  
\bibitem[\protect\citeauthoryear{Bruzual and Charlot}{2003}]{bc03}
  \reference{Bruzual~G. and Charlot~S.}
  {\journal{\mnras}{\bf 344}{1000}{2003}}

\bibitem[\protect\citeauthoryear{Burenin \etal}{2007}]{400d}
  \reference{Burenin \etal}[Burenin~R., Vikhlinin~A., Hornstrup~A., Ebeling~H., Quintana~H. \etal]
  {\journal{\apjs}{\bf 172}{561}{2007}}
  
\bibitem[\protect\citeauthoryear{Burenin}{2017}]{br17}
  \reference{Burenin~R.~A.}
  {\journal{\astl}{\bf 43}{507}{2017}}
  
\bibitem[\protect\citeauthoryear{Burenin \etal}{2018}]{br2018hz}
  \reference{Burenin \etal}[Burenin~R.~A., Bikmaev~I.~F., Khamitov~I.~M., Zaznobin~I.~A., Khorunzhev~G.~A. \etal]
  {\journal{\astl}{\bf 44}{297}{2018}}

\bibitem[\protect\citeauthoryear{Burenin \etal}{2021}]{br22hz}
  \reference{Burenin \etal}[Burenin~R.~A., Bikmaev~I.~F., Gilfanov~M.~R., Grokhovskaya~A.~A., Dodonov~S.~N. \etal]
  {\journal{\astl}{\bf 47}{443}{2021}}
  
\bibitem[\protect\citeauthoryear{Burenin}{2022a}]{br22}
  \reference{Burenin~R.~A.}
  {\journal{\astl}{\bf 48}{153}{2022}}
  
\bibitem[\protect\citeauthoryear{Burenin \etal}{2022b}]{br23} 
  \reference{Burenin~R.~A. \etal}[Burenin~R.~A., Zaznobin~I.~A., Medvedev~P.~S., Gilfanov~M.~R., Kotov~S.~S. \etal]
  {\journal{\astl}{\bf 48}{702}{2022}} 
  
\bibitem[\protect\citeauthoryear{Chambers \etal}{2016}]{ps1}
  \reference{Chambers \etal}[Chambers~K.~C., Magnier~E.~A., Metcalfe~N., Flewelling~H.~A., Huber~M.~E. \etal]
  {\journal{arXiv:1612.05560}{}{}{2016}}

\bibitem[\protect\citeauthoryear{Desprez \etal}{2020}]{euclid}
  \reference{Desprez \etal}[Desprez~G., Paltani~S., Coupon~J., Almosallam~I., Alvarez-Ayllon~A. \etal]
  {\journal{\aap}{\bf 644}{A31}{2020}}

\bibitem[\protect\citeauthoryear{Dey \etal}{2019}]{desi}
  \reference{Dey \etal}[Dey~A., Schlegel~D.~J., Lang~D., Blum~R., Burleigh~K., \etal]
  {\journal{\aj}{\bf 157}{id. 168}{2019}}

\bibitem[\protect\citeauthoryear{Dickey and Lockman}{1990}]{nh}
  \reference{Dickey~J.~M. and Lockman~F.~J.}
  {\journal{\araa}{\bf 28}{215}{1990}}

\bibitem[\protect\citeauthoryear{Ebeling \etal}{2001}]{macs}
  \reference{Ebeling~H., Edge~A.~C., Henry~J.~P.}
  {\journal{\apj}{\bf 533}{668}{2019}}

\bibitem[\protect\citeauthoryear{Hernández-Lang \etal}{2022}]{madpsz}
  \reference{Hernández-Lang \etal}[Hernández-Lang~D., Mohr~J.~J., Klein~M., Grandis~S., Melin~J.-B. \etal]
  {\journal{\mnras}{\bf 517}{4355}{2022}}

\bibitem[\protect\citeauthoryear{Hilton \etal}{2018}]{zcluster}
  \reference{Hilton \etal}[Hilton~M., Hasselfield~M., Sifón~C., Battaglia~N., Aiola~S. \etal]
  {\journal{\apjs}{\bf 235}{20}{2018}}   
  
\bibitem[\protect\citeauthoryear{Hilton \etal}{2021}]{act}
  \reference{Hilton \etal}[Hilton~M., Sifón~C., Naess~S., Madhavacheril~M., Oguri~M. \etal]
  {\journal{\apjs}{\bf 253}{3}{2021}}

\bibitem[\protect\citeauthoryear{Khamitov \etal}{2020}]{rttmos20}
  \reference{Khamitov \etal}[Khamitov~I.~M., Bikmaev~I.~F., Burenin~R.~A., Glushkov~M.~V., Melnikov~S.~S. \etal]
  {\journal{\astl}{\bf 46}{1}{2020}}

\bibitem[\protect\citeauthoryear{Klein \etal}{2018}]{mcmf} 
  \reference{Klein \etal}[Klein~M., Mohr~J.~J., Desai~S., Israel~H., Allam~S., \etal]
  {\journal{\mnras}{\bf 474}{3324}{2018}}  

\bibitem[\protect\citeauthoryear{Kopylova and Kopylov}{2006}]{Kopylova}
  \reference{Kopylova~F.~G. and Kopylov~A.~I.} 
  {\journal{\astl}{\bf 32}{95}{2006}}

\bibitem[\protect\citeauthoryear{Lin \etal}{2004}]{lin} 
  \reference{Lin~Y.-T., Mohr~J.~J., Stanford~S.~A.}
  {\journal{\apj}{\bf 610}{745}{2004}}

\bibitem[\protect\citeauthoryear{Meshcheryakov \etal}{2015}]{mesh}
  \reference{Meshcheryakov~A.~V., Glazkova~V.~V., Gerasimov~S.~V., Burenin~R.~A., Khorunzhev~G.~A.}
  {\journal{\astl}{\bf 41}{307}{2015}}

\bibitem[\protect\citeauthoryear{Mullis \etal}{2003}]{160d}
  \reference{Mullis \etal}[Mullis~C.~R., McNamara~B.~R., Quintana~H., Vikhlinin~A., Henry~J.~P., \etal]
  {\journal{\apj}{\bf 594}{154}{2003}}

\bibitem[\protect\citeauthoryear{Pavlinsky \etal}{2021}]{art}
  \reference{Pavlinsky \etal}[Pavlinsky~M., Tkachenko~A., Levin~V., Alexandrovich~N., Arefiev~V. \etal.]
  {\journal{\aap}{{\bf 650}}{18}{2021}} 

\bibitem[\protect\citeauthoryear{Piffaretti \etal}{2011}]{mcxc} 
  \reference{Piffaretti~R., Arnaud~M., Pratt~G.~W., Pointecouteau~E., Melin~J.-B.}
  {\journal{\aap}{\bf 534}{A109}{2011}} 

\bibitem[\protect\citeauthoryear{Planck Collaboration}{2014a}]{PSZcosm13}
  \reference{Planck Collaboration}[Planck 2013 Results XX: Ade~P.~A.~R., Aghanim~N., Armitage-Caplan~C. \etal]
  {\journal{\aap}{\bf 571}{A20}{2014a}}
  
\bibitem[\protect\citeauthoryear{Planck Collaboration}{2014b}]{PSZ1}
  \reference{Planck Collaboration}[Planck 2013 Results XXIX: Ade~P.~A.~R., Aghanim~N., Armitage-Caplan~C. \etal] 
  {\journal{\aap}{\bf 571}{A29}{2014b}}
  
\bibitem[\protect\citeauthoryear{Planck Collaboration}{2015a}]{PSZ_RTT150}
  \reference{Planck Collaboration}[Planck Intemediate Results XXVI: Ade~P.~A.~R., Aghanim~N., Arnaud~M. \etal] 
  {\journal{\aap}{\bf 582}{A29}{2015a}}
  
\bibitem[\protect\citeauthoryear{Planck Collaboration}{2015b}]{PSZ1Addendum}
  \reference{Planck Collaboration}[Planck 2013 Results XXXII: Ade~P.~A.~R., Aghanim~N., Armitage-Caplan~C. \etal] 
  {\journal{\aap}{\bf 581}{A14}{2015b}}  

\bibitem[\protect\citeauthoryear{Planck Collaboration}{2016a}]{PSZ2}
  \reference{Planck Collaboration}[Planck Collaboration, Ade~P.~A.~R., Aghanim~N., Arnaud~M. \etal]
  {\journal{\aap}{\bf 594}{38}{2016a}}
  
\bibitem[\protect\citeauthoryear{Planck Collaboration}{2016b}]{PSZ2cosm}
  \reference{Planck Collaboration}[Planck 2015 Results XXIV: Ade~P.~A.~R., Aghanim~N., Arnaud~M. \etal]
    {\journal{\aap}{\bf 594}{A24}{2016b}}
  
\bibitem[\protect\citeauthoryear{Predehl \etal}{2021}]{pred21} 
  \reference{Predehl \etal}[Predehl~P., Andritschke~R., Arefiev~V., Babyshkin~V., Batanov~O., Becker~M. \etal]
  {\journal{\aap}{\bf 647}{16}{2021}}

\bibitem[\protect\citeauthoryear{Rykoff \etal}{2014}]{redmapper} 
  \reference{Rykoff \etal}[Rykoff~E.~S., Rozo~E., Busha~M.~T., Cunha~C.~E., finoguenov~A. \etal]
  {\journal{\apj}{\bf 785}{33}{2014}}

\bibitem[\protect\citeauthoryear{Sarazin}{1986}]{sarazin}
  \reference{Sarazin~C.~L.}
  {\journal{Rev. Modern Phys.}{\bf 58}{1}{1986}}

\bibitem[\protect\citeauthoryear{Schmidt \etal}{2020}]{lsst}
  \reference{Schmidt \etal}[Schmidt~S.~J., Malz~A.~I., Soo~J.~Y.~H., Almosallam~I.~A., Brescia~M., \etal]
  {\journal{\mnras}{\bf 499}{1587}{2020}}

\bibitem[\protect\citeauthoryear{SDSS Collaboration}{2016}]{sdssdr13} 
  \reference{SDSS Collaboration: Ahumada~R., Allende Prieto~C., Almeida~A., Anders~F., Anders~S.~F.\etal}
  {\journal{\apjs}{\bf 233}{2}{2016}}
  
\bibitem[\protect\citeauthoryear{SDSS Collaboration}{2020}]{sdssdr16} 
  \reference{SDSS Collaboration: Ahumada~R., Allende Prieto~C., Almeida~A., Anders~F., Anders~S.~F.\etal}
  {\journal{\apjs}{\bf 249}{3}{2020}}

\bibitem[\protect\citeauthoryear{Streblyanska \etal}{2018}]{Stre}
  \reference{Streblyanska \etal}[Streblyanska~A., Barrena~R., Rubiño-Martín~J.~A., van~der~Burg~R.~F., Aghanim~N. \etal]
  {\journal{\aap}{\bf 617}{A71}{2018}}

\bibitem[\protect\citeauthoryear{Sunyaev \etal}{2021}]{srg} 
  \reference{Sunyaev \etal}[Sunyaev~R., Arefiev~V., Babyshkin~V., Bogomolov~A., Borisov~K. \etal]
  {\journal{\aap}{{\bf Special issue “The Early Data Release of eROSITA and Mikhail Pavlinsky ART-XC on the SRG mission”}}{}{2021}}
  
\bibitem[\protect\citeauthoryear{Vikhlinin \etal}{2009a}]{av09xr}
  \reference{Vikhlinin \etal}[Vikhlinin~A., Burenin~R.~A., Ebeling~H., Forman~W.~R., Hornstrup~A. \etal]
  {\journal{\apj}{\bf 692}{1033}{2009}}
  
\bibitem[\protect\citeauthoryear{Vikhlinin \etal}{2009b}]{av09cosm}
  \reference{Vikhlinin \etal}[Vikhlinin~A., Kravtsov~A.~V., Burenin~R.~A., Forman~W.~R., Hornstrup~A. \etal]
  {\journal{\apj}{\bf 692}{1060}{2009}}  

\bibitem[\protect\citeauthoryear{Vorobyev \etal}{2016}]{vorobyev16}
  \reference{Vorobyev \etal}[Vorobyev~V.~S., Burenin~R.~A., Bikmaev~I.~F., Khamitov~I.~M., Dodonov~S.~N. \etal]
  {\journal{\astl}{\bf 42}{63}{2016}}
  
\bibitem[\protect\citeauthoryear{Wen \etal}{2012}]{whl}
  \reference{Wen~Z.~L., Han~J.~L., Liu~F.~S.}
  {\journal{\apjs}{\bf 199}{12}{2012}}  
  
\bibitem[\protect\citeauthoryear{Wright \etal}{2010}]{wright10} 
  \reference{Wright \etal}[Wright~E.~L., Eisenhardt~P.~R.~M., Mainzer~A.~K., Ressler~M.~E., Cutri~R.~M. \etal]
  {\journal{\aj}{\bf 140}{1868}{2010}}  
  
\bibitem[\protect\citeauthoryear{Zaznobin \etal}{2019}]{zazn19}
  \reference{Zaznobin \etal}[Zaznobin~I.~A., Burenin~R.~A., Bikmaev~I.~F., Khamitov~I.~M., Khorunzhev~G.~A. \etal]
  {\journal{\astl}{\bf 45}{49}{2019}}
  
\bibitem[\protect\citeauthoryear{Zaznobin \etal}{2020}]{zazn20}
  \reference{Zaznobin \etal}[Zaznobin~I.~A., Burenin~R.~A., Bikmaev~I.~F., Khamitov~I.~M., Khorunzhev~G.~A. \etal]
  {\journal{\astl}{\bf 46}{79}{2020}}
  
\bibitem[\protect\citeauthoryear{Zaznobin \etal}{2021a}]{zazn21}
  \reference{Zaznobin \etal}[Zaznobin~I.~A., Burenin~R.~A., Bikmaev~I.~F., Khamitov~I.~M., Khorunzhev~G.~A. \etal]
  {\journal{\astl}{\bf 47}{61}{2021}}

\bibitem[\protect\citeauthoryear{Zaznobin \etal}{2021b}]{zazn21lh}
  \reference{Zaznobin \etal}[Zaznobin~I.~A., Burenin~R.~A., Lyapin~A.~R., Khorunzhev~G.~A., Afanasiev~V.~L. \etal]
  {\journal{\astl}{\bf 47}{141}{2021}}

\bibitem[\protect\citeauthoryear{Zou \etal}{2019}]{zou19}
  \reference{Zou \etal}[Zou~H., Gao~J., Zhou~X., Kong~X.]
  {\journal{\apjs}{\bf 242}{8}{2019}} 
  
\bibitem[\protect\citeauthoryear{Zou \etal}{2022}]{zou}
  \reference{Zou \etal}[Zou~H., Sui~J., Xue~S., Zhou~X., Ma~J. \etal]
  {\journal{Research in Astronomy and Astrophysics}{\bf 22}{id.065001}{2022}}
  
\end{thebibliography}
\end{document}